\documentclass[twocolumn,journal]{revtex4-1}
\usepackage[T1]{fontenc}
\usepackage{amsmath}
\usepackage{amssymb}
\usepackage{graphicx}
\usepackage[unicode=true,
 bookmarks=true,bookmarksnumbered=true,bookmarksopen=true,bookmarksopenlevel=1,
 breaklinks=false,pdfborder={0 0 0},backref=false,colorlinks=false]
 {hyperref}
\hypersetup{pdftitle={Your Title},
 pdfauthor={Your Name},
 pdfpagelayout=OneColumn, pdfnewwindow=true, pdfstartview=XYZ, plainpages=false}
\usepackage{breakurl}

\makeatletter
\usepackage[caption=false,font=footnotesize]{subfig}

\makeatother

\begin{document}

\title{A high-resolution programmable Vernier delay generator based on carry
chains in FPGA}

\author{Ke~Cui,$^{1}$ Xiangyu~Li,$^{2\text{,}\ast}$ and~Rihong~Zhu$^{1}$}

\affiliation{$^{1}$the MIIT Key Laboratory of Advanced Solid Laser, Nanjing University
of Science and Technology, \#200 Xiaolingwei, Nanjing, Jiangsu, China}

\affiliation{$^{2}$the School of Computer Science and Engineering, Nanjing University
of Science and Technology, \#200 Xiaolingwei, Nanjing, Jiangsu, China}

\email{Electronic mail: wl.njust.edu.cn}

\begin{abstract}
This paper presents an architecture of high-resolution delay generator
implemented in a single field programmable gate array (FPGA) chip
by exploiting the method of utilizing dedicated carry chains. It serves
as the core component in various physical instruments. The proposed
delay generator contains the coarse delay step and the fine delay
step to guarantee both large dynamic range and high resolution. The
carry chains are organized in the Vernier delay loop style to fulfill
the fine delay step with high precision and high linearity. The delay
generator was implemented in the EP3SE110F1152I3 Stratix III device
from Altera on a self-designed test board. Test results show the obtained
resolution is 38.6 ps, and the differential nonlinearity (DNL)\,\textbackslash{}\,integral
nonlinearity (INL) is in the range of (-0.18 least significant bit
(LSB), 0.24 LSB)\,\textbackslash{}\,(-0.02 LSB, 0.01 LSB) under
the nominal supply voltage of 1100 mV and environmental temperature
of 20 $^{\circ}$C. The delay generator is also rather resource cost
efficient which uses only 668 LUTs and 146 registers in total.
\end{abstract}


\maketitle

\section{Introduction}

Highly accurate programmable delay generator is used to generate an
expected time interval between two successive pulses. It is manifested
as core component in many physical instruments, for example in the
positron emission tomography (PET) system \cite{YLiu2003}, in the
STACEE gamma-ray astrophysics experiment \cite{JPMartin2000} and
in the conflict detector applications for multi-synchronous systems
\cite{RGinosar2011}. Besides, precise delay generator is very beneficial
for any timing experiments and tests \cite{HUJang-RSI},\cite{JKalisz-RSI}.

Generally speaking delay generators can be divided into two classifications:
absolute delay generators and relative generators. Absolute generators
directly produce a delayed output signal relative to the input trigger
signal which usually contains an inevitable intrinsic fixed delay
making the delay result less accurate. Relative generators produce
two new delay outputs between which the time interval equals the programmed
delay amount after an effective detection of the input trigger signal.
Relative generators can achieve much higher accuracy and are the main
concerns in this paper. However absolute delay generators may be necessarily
required under some circumstances, even so relative delay generators
can also be applied to these cases by making proper adjustment of
the utilization manner of the output delay signals. If this is applicable,
the leading delay output of the relative delay generator acts as the
new time base or effective trigger signal instead of the actual input
trigger signal. 

In this paper we focus on the designing of precise delay generators
by utilizing the field programmable gate array (FPGA) chip, which
is widely adopted in many scientific instruments mainly benefiting
from its real-time signal controlling and processing ability and flexible
reconfigurability. Y. Song proposed an FPGA-based delay generator
with 65 ps resolution and 400 ps RMS jitter. However the delay was
not fully generated by a single FPGA chip and needed to be calibrated
by an extra fine delay chip AD9501 \cite{YSong2011}. R.Giordano proposed
a high-resolution synthesizable delay line in a Xilinx Kintex-7 FPGA
with 11 ps resolution and 3.0 ns delay range \cite{RGiordano2015}.
It opened up a new way to build high-performance delay line without
any custom design intervene.

The dedicated carry chains, which are especially designed to implement
fast arithmetic functions such as adders, counters and comparators,
are important components of modern FPGAs. Since the propagation delay
between two consecutive basic elements along the carry chain is rather
small, it has the promise to fulfill the delay generating task accurately.
Many time-to-digital (TDC) \cite{KeCui_TDC2017}-\nocite{JYWon_TDC2016}\cite{WangYG_TDC2016}
converter designs, which in function can be seen as the reversion
of the delay generator, have been deeply exploited by utilizing the
carry chain. The main contribution of this paper is to extend the
carry-chain-based structure to the designing of high performance delay
generator and to demonstrate the effectiveness of this method. The
carry chain is organized in the Vernier loop style to generate the
fine delay resolution at a level of tens of ps which will be fully
illustrated in the following sections.

\section{design method}

\begin{figure*}[tbh]
\centering\includegraphics[width=0.9\textwidth]{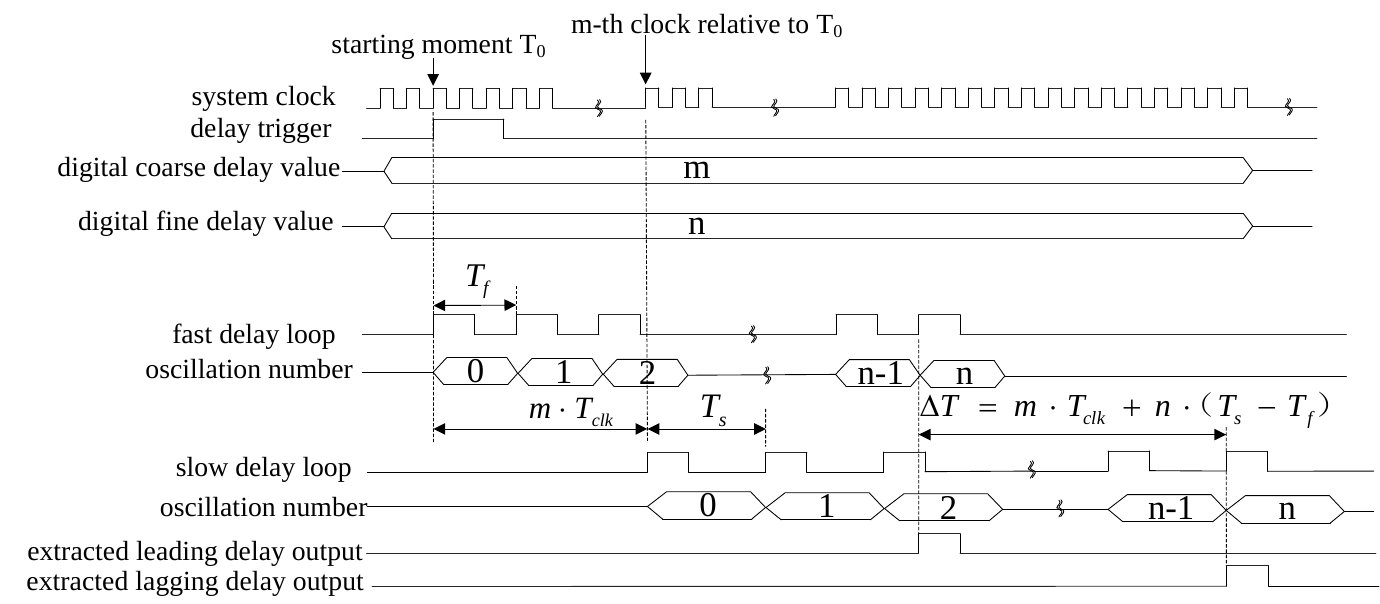}\caption{Timing diagram of the proposed delay generator.}
\end{figure*}

Our goal is to exploit the possibility of utilizing the carry chain
to construct the two delay lines working in the Vernier mode. Considering
that any mismatch between the the delay elements of the Vernier delay
line may result in linearity error, an ideal circuit implementation
should require all delay elements to distribute equally and symmetrically
in the FPGA chip to yield the same delay. However it is impossible
for such a strict requirement to be met due to the uncontrollable
fabrication parameter variations during the CMOS processes. To combat
this problem, we adopt the technique of folding the long delay line
into shorter loop to alleviate the risk of nonlinearity \cite{JYu2010}.
The oscillation periods of the slow and fast delay loops organized
in the Vernier type are denoted as $T_{s}$ and $T_{f}$ respectively.
After each time of oscillation a fixed delay difference $r_{f}=T_{s}-T_{f}$
is produced. So a delay interval can be digitally programmed by controlling
the overall oscillation number $n$. 

It deserves noticing that the dead time for this type of delay generator
is at least $n\cdot T_{s}$ which turns out to be very large for long
delay interval. In order to cover a large dynamic range, a two-step
delay generation method is adopted. The first step is the coarse delay
generation process and the second step is the fine delay generation
process. We use the combination $(m,n)$ and $(r_{c},r_{f})$ to denote
the programmable delay values and resolutions respectively, where
$m,\thinspace r_{c}$ are the setting parameters for the coarse delay
step and $n,\thinspace r_{f}$ are the setting parameters for the
fine delay step. Usually the $(r_{c},\thinspace r_{f})$ is chosen
to satisfy: $r_{c}>10r_{f}$ to ensure both fine resolution and large
dynamic range. If a delay interval of $\varDelta T$ is expected,
the corresponding parameters are calculated as: $m=\left\lfloor \varDelta T/r_{c}\right\rfloor $,
$n=\left\lfloor (\varDelta T-m\cdot r_{c})/r_{f}\right\rfloor $,
where $\left\lfloor x\right\rfloor $ denotes the largest integer
less than or equal to $x$. 

Fig.1 shows the timing diagram of the proposed delay generator. At
the starting moment T$_{0}$, the delay generator latches the predefined
digital delay values and the delay trigger signal and then simultaneously
generates two delay pulses. One of the delay pulses is fed directly
to the fast delay loop while the other is fed to the slow delay loop
after being delayed by $m\cdot r_{c}$. Since $r_{c}$ can be chosen
large enough, the coarse delay step can be fulfilled by a counter.
After the delay pulses enter the Vernier delay loop, the fine delay
step begins to work by setting up two oscillations. It is obvious
that the oscillation in the fast delay loop begins earlier as long
as $m$ is nonzero. The delay loops are labeled each with the output
of a fine counter to record the oscillation number. The oscillation
pulse along each delay loop is connected to the clock port of its
corresponding fine counter which runs upward. When the counter value
equals $n$, the corresponding oscillation pulse is extracted as the
delay signal. The delay signal from the fast delay line acts as the
leading output while the one from the slow delay line acts as the
lagging output. So the delay interval generated between the two delay
signals is: 
\begin{equation}
\begin{array}{c}
\varDelta T=\varDelta T_{c}+\varDelta T_{f}=m\cdot r_{c}+n\cdot r_{f}\\
=m\cdot T_{clk}+n\cdot(T_{s}-T_{f})
\end{array}
\end{equation}

\section{circuit implementation}

\begin{figure*}[tbh]
\centering\includegraphics[width=0.9\textwidth]{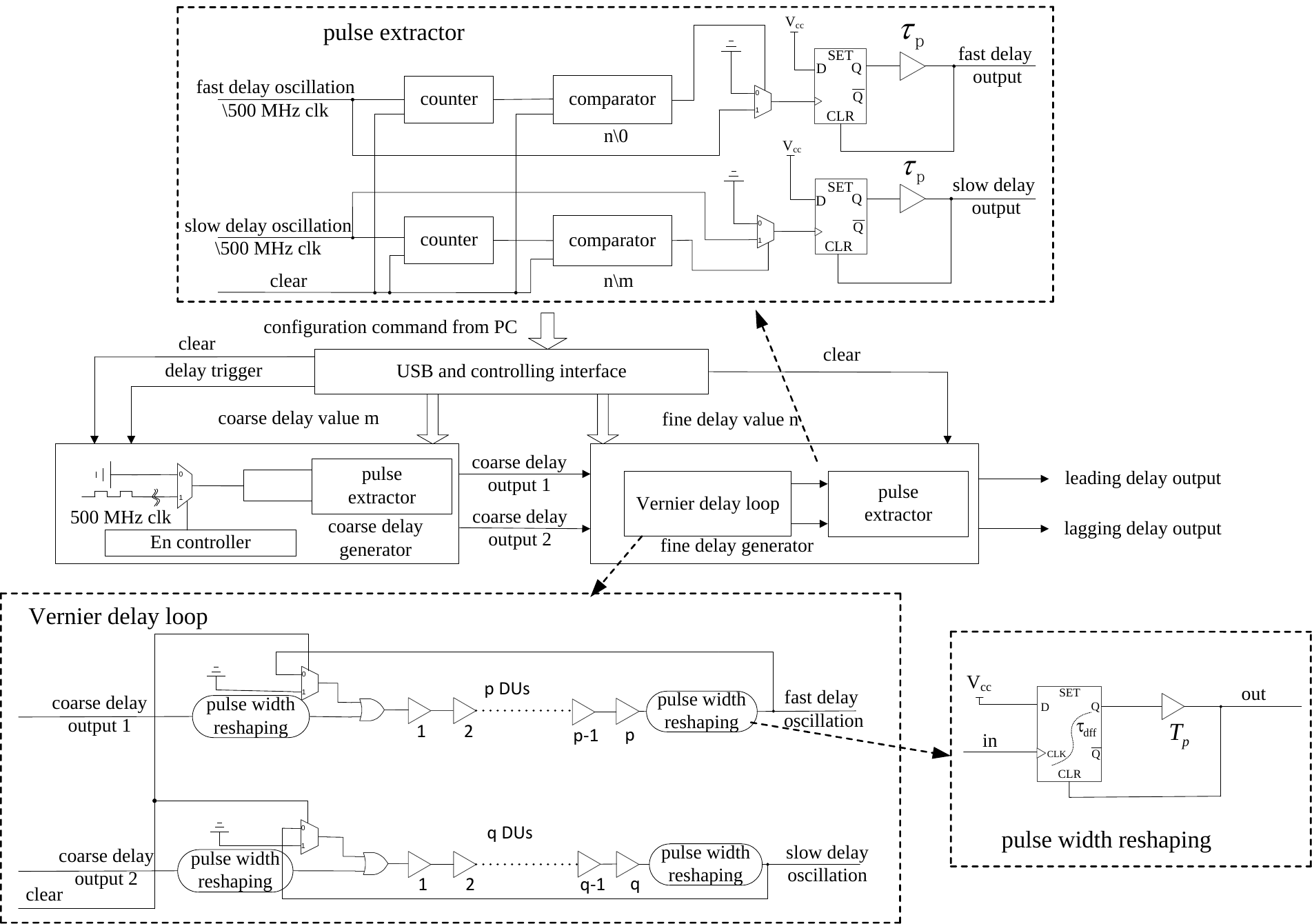}\caption{Implementation circuit of the proposed delay generator.}
\end{figure*}

The global architecture of the proposed delay generator is depicted
in Fig.2. It mainly contains three parts: the USB and controlling
interface, the coarse delay generator and the fine delay generator.
The USB and controlling interface is the controlling core of the delay
generator. It is in charge of receiving digitally programmable delay
values from PC via USB bus, producing the delay trigger and clearing
the status after each time of working. 

The coarse delay generator is mainly composed of one pulse extractor
module whose input ports are fed by the 500 MHz clock signal (denoted
as system clock in Fig.1) resulting in the coarse delay resolution
of 2 ns which is firstly multiplexed by an En controller module. The
En controller module turns to high level after it detects the positive
status of the delay trigger at the starting moment T$_{0}$. According
to the functional principle of the pulse extractor, whenever the 500
MHz is selected to pass through the multiplexer, the two counters
in the pulse extractor are triggered to run upward. Two delay outputs
are generated when the counters equal 0 and $m$ respectively. The
first generated output is fed to the fast delay loop while the other
to the slow delay loop, by this way generating the coarse time interval
$m\cdot T_{clk}$ before the fine delay step functions.

The pulse extractor module is designed to locate and extract the expected
pulse from the corresponding oscillation sequence. The pulse to be
extracted is predefined by the delay value $m$ or $n$. This module
contains two counters, two comparators and two D-type flip-flops as
shown in Fig.2. The two counters run upward driven by the oscillation
sequence from which the expected pulse is to be extracted. The two
counters are both driven by the 500 MHz system clock in the coarse
delay generator, while respectively by the fast and slow oscillation
pulses in the fine delay generator. The counter's output is connected
to its following comparator whose comparison counterpart is one of
the customized preset values of 0, $m$ and $n$. The two comparators's
preset values are 0 and $m$ respectively in the coarse delay generator
while both $n$ in the fine delay generator. When the counter value
equals its comparison counterpart, the comparator outputs a rising
edge to activate the D-type flip-flop to generate a delay output.
In the special case of the preset value of 0, since it equals the
default value of the counter making the comparator to output a constant
positive level, the multiplexer is always switched to the 1-th channel.
That means the first pulse of the corresponding oscillation signal
will be extracted which exactly has zero added delay. A fixed delay
buffer is inserted between the Q port and the CLR port of the flip-flop
to constrain the pulse width to a proper value. 

The fine delay generator contains a Vernier delay loop and another
pulse extractor. To facilitate the explanation, the smallest and indivisible
basic delay cell along the carry chain is denoted as delay unit (DU)
throughout this paper. The Vernier delay loop contains two delay loops
constituted of some amount of DUs along the dedicated carry chain
in the FPGA (the maximal amount is set as 32 in our example design).
Each loop consists of one delay element, one fixed delay buffer, one
pulse width reshaping module, one multiplexer and one OR gate. The
pulse width reshaping module is designed to constrain and stabilize
the positive duration $T_{p}$ of the oscillation pulse which should
be less than $T_{f}$. It uses a D flip-flop followed by a delay buffer
to fulfill the reshaping function. In our example design, $T_{p}=2$
ns and $T_{f}\thickapprox T_{s}=4$ ns.

\begin{figure*}[tbh]
\centering\includegraphics[width=0.9\textwidth]{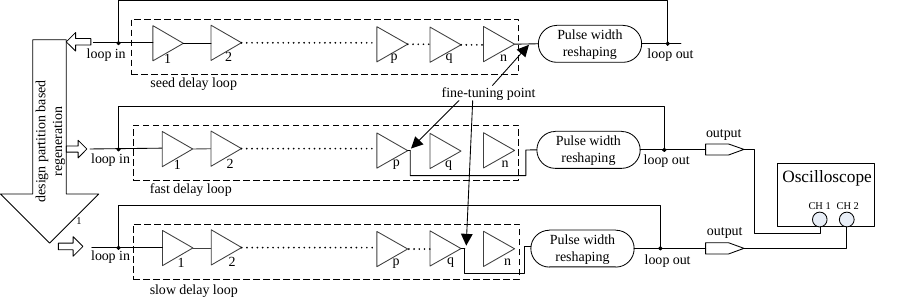}\caption{Design details of the Vernier delay loop.}
\end{figure*}

\begin{figure}[tbh]
\centering\includegraphics[width=0.9\columnwidth]{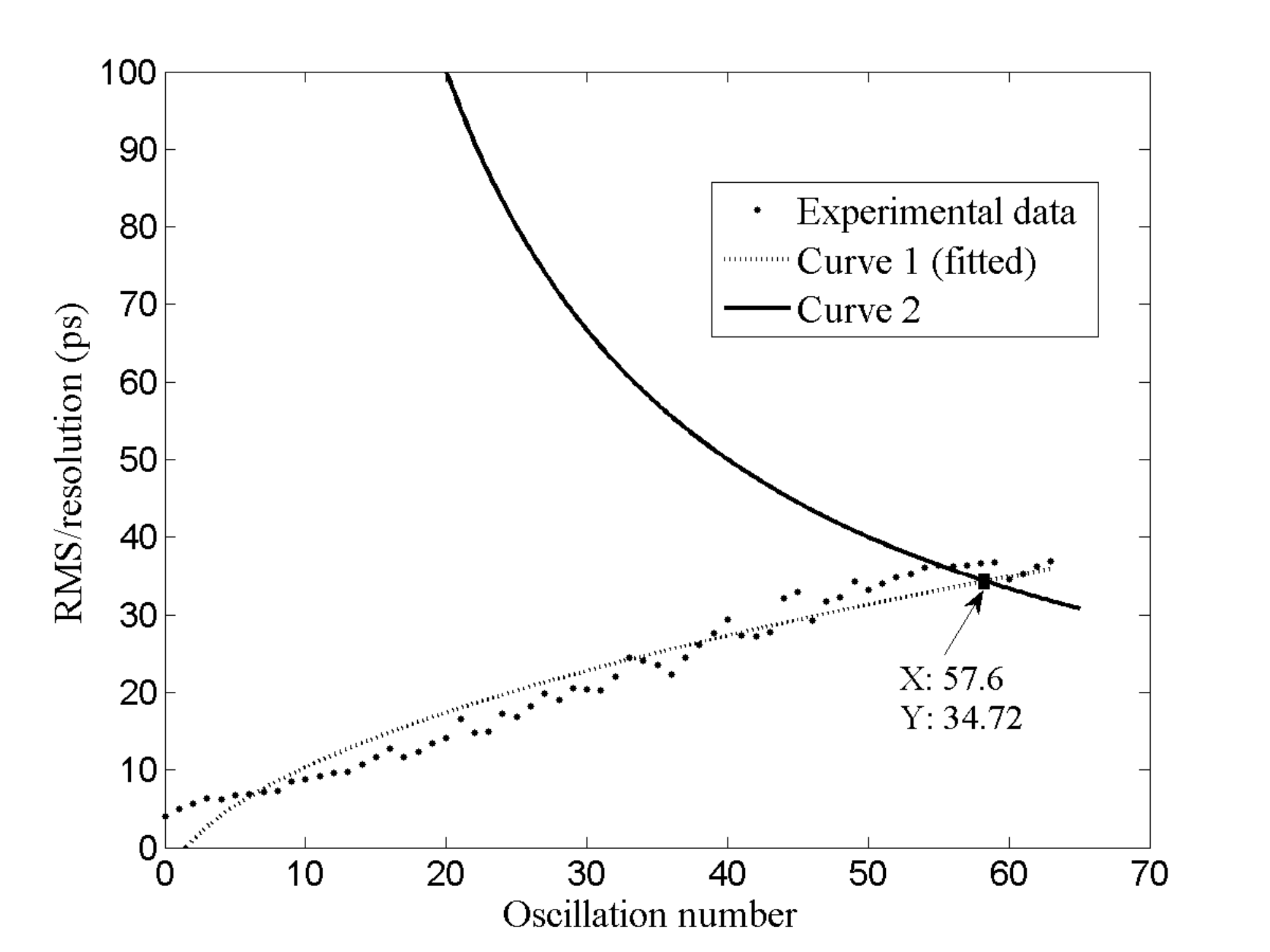}\caption{Resolution versus the fine oscillation number.}
\end{figure}

Obviously construing the two delay loops with slightly different oscillation
periods is the most challenging task, and in order to overcome this
design difficulty we propose an efficient two-step construction method.
It works as follows: in the first step, a delay loop of the structure
as shown in Fig.2 is built in a standalone Quartus project and then
exported out as the seed loop. In the second step, a new Quartus project
is set up and two delay loops are built from the same seed by applying
importation operation. The two-step construction method guarantees
that the structures of the two delay loops are identical which should
ideally lead to an equal oscillation period. However any delay mismatch
of the DUs in the two delay loops will break the equation condition
and make the oscillation periods different from each other in a random
manner. Fortunately according to our building experience, this period
difference (or resolution) is always small enough such that it can
provide the basis to further finely tune the two delay loops to obtain
a desired resolution. The fine-tuning process is applied at the fine-tuning
point which is actually the wire connection at the end of each delay
loop as shown in Fig.3. The oscillation signals along the two delay
loops are brought out from the two fine-tuning points and connected
to an external oscilloscope. By observing the relative oscillation
periods from the oscilloscope, the resolution of the current delay
loops configuration can be deduced. To finely tune the resolution,
the following actions are applied: pick up the delay loop that corresponds
to the longer oscillation period, disconnect the delay loop at the
fine-tuning point, shorten its length by 1 DU, and reconnect it to
form the new shorter delay loop. Those actions are iteratively performed
until a desired resolution is found. It should be noticed that the
regeneration process of the fine-tuning point would change the capacitance
condition of the final DU of the delay loops and cause an extra delay.
However this effect is not harmful to the fine-tuning process which
is an iterative searching process containing many DU number combination
possibilities. Some similar works on manually trimming delay path
circuits to obtain a desired resolution can be found in \cite{JKalisz_TDC1997},\cite{HaiWang_TDC2013}.
In our example design, we realized the mentioned operation using the
engineering change orders (ECO) tool provided by the Quartus II software.
Our design experience verifies that a wide resolution range of (10
ps, 100 ps) can be constructed. 

Small period difference improves the fine delay resolution but raises
the risk of larger RMS jitter. This can be illustrated by the following
formulas which expresses the proportional relationship between the
RMS jitter of the delay interval $\varDelta T$ and the square root
of the oscillation time \cite{AAAbidi2006}:

\begin{equation}
\sigma_{\vartriangle T}=p\sqrt{n\cdot T_{s}}=p\sqrt{\frac{\varDelta T-m\cdot r_{c}}{r_{f}}\cdot T_{s}}\label{eq:rms}
\end{equation}

where $p$ is a circuit-dependent factor. Here only the fine delay
RMS jitter is considered because the coarse delay RMS jitter can be
stabilized by the PLL's feedback loop and is therefore omitted. Equation
(\ref{eq:rms}) means the RMS jitter is inversely proportional to
square root of the fine resolution $r_{f}$ which restricts its smallest
choice. We experimentally tested the RMS jitter versus the oscillation
number $n$ and draw the fitted curve in Fig.4 (curve 1). An assistant
curve 2 which is modeled by equation $r_{f}=T_{clk}/n$ representing
the resolution versus $n$ is also given. These two curves generate
a crossing point which has the resolution value of about 35 ps. This
point represents the lower resolution bound that guarantees the corresponding
RMS jitter not larger than 1 LSB. That means a careful performance
tradeoff between the resolution and RMS jitter should be made, and
in our example design we choose the resolution of 35 ps as the construction
target. 

\begin{figure}[tbh]
\centering\includegraphics[width=0.9\columnwidth]{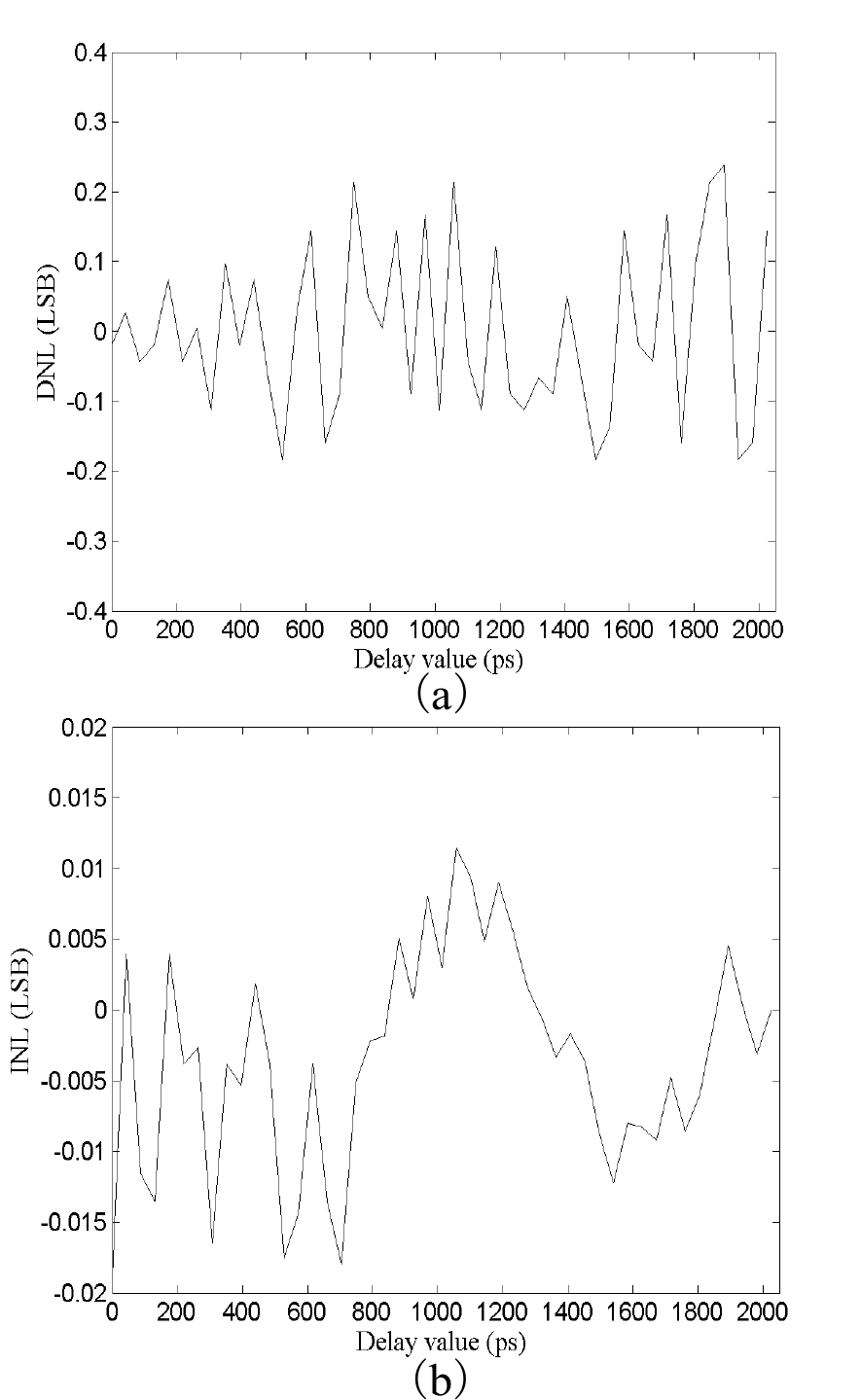}\caption{Measured (a) DNL and (b) INL of the delay generator under the nominal
supply voltage of 1100 mV and environmental temperature of 20 $^{\circ}$C.}
\end{figure}

\begin{figure}[tbh]
\centering\includegraphics[width=0.9\columnwidth]{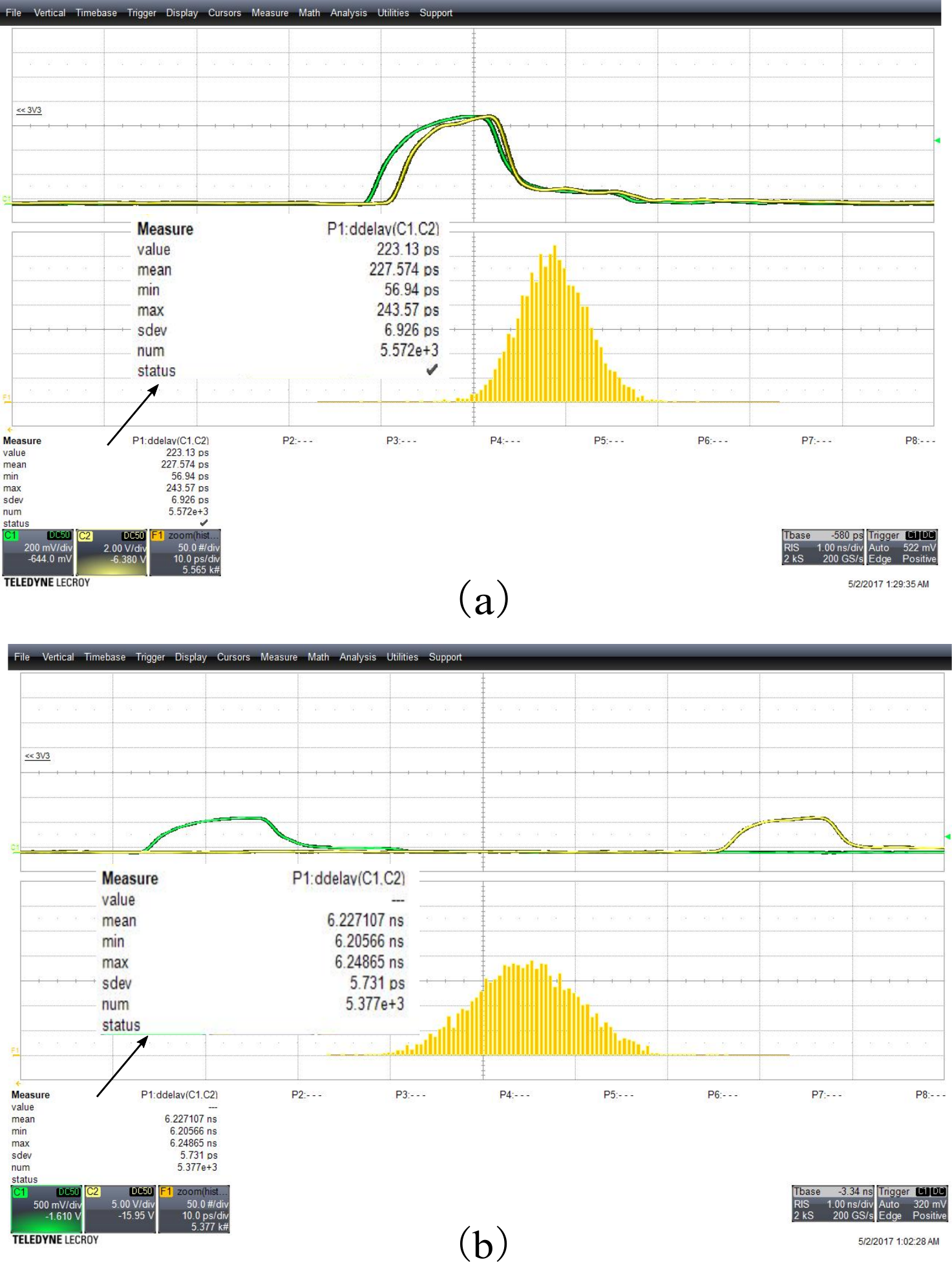}\caption{Waveform and histogram of the delay generator captured by the 740Zi
Lecroy oscilloscope when the preset delay values are (a) 232 ps and
(b) 6232 ps.}
\end{figure}

\section{Test results}

\begin{figure}[tbh]
\centering\includegraphics[width=1\columnwidth]{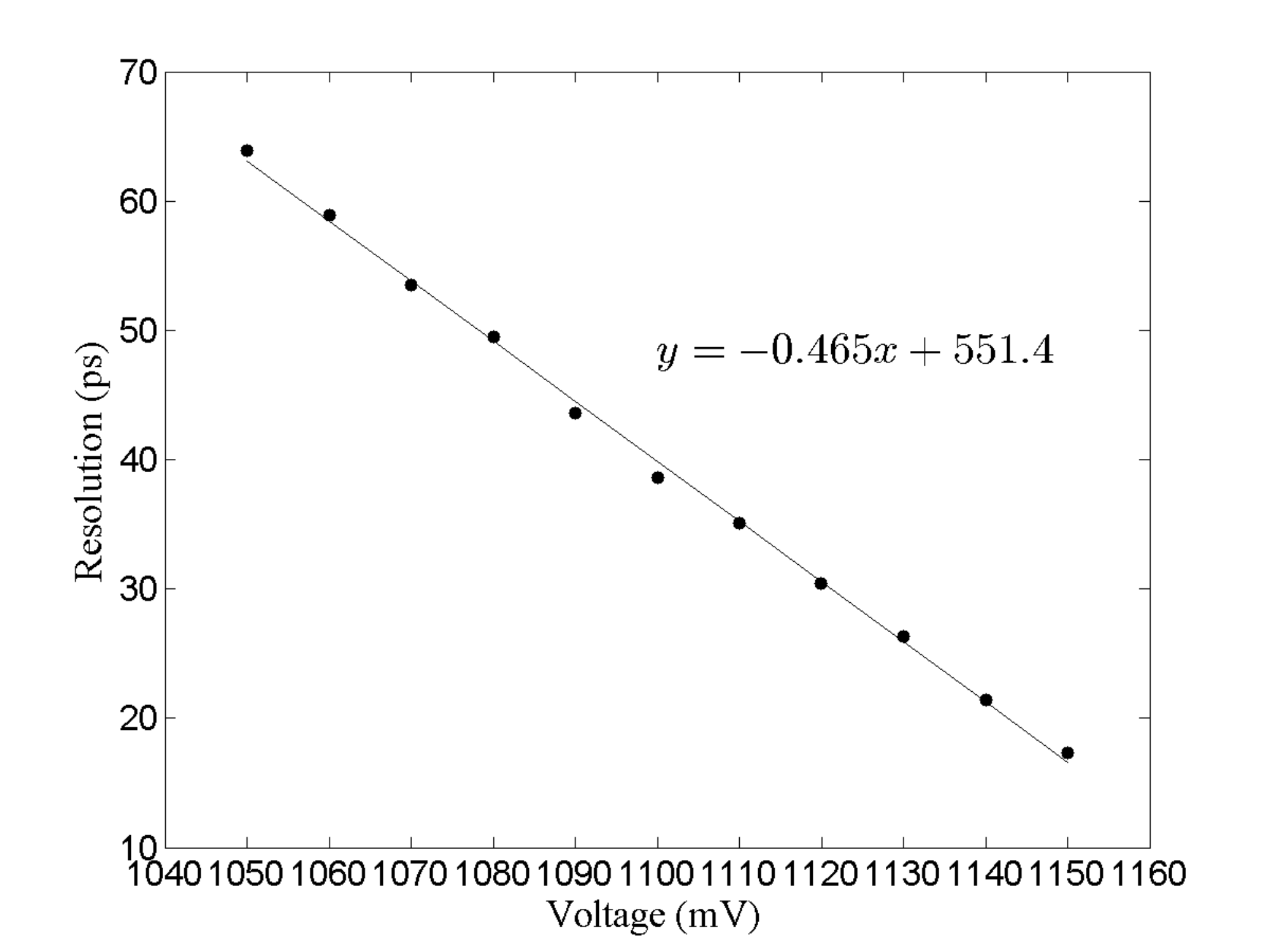}

\caption{Resolution versus the core voltage of the FPGA chip.}
\end{figure}

\begin{figure}[tbh]
\centering\includegraphics[width=0.9\columnwidth]{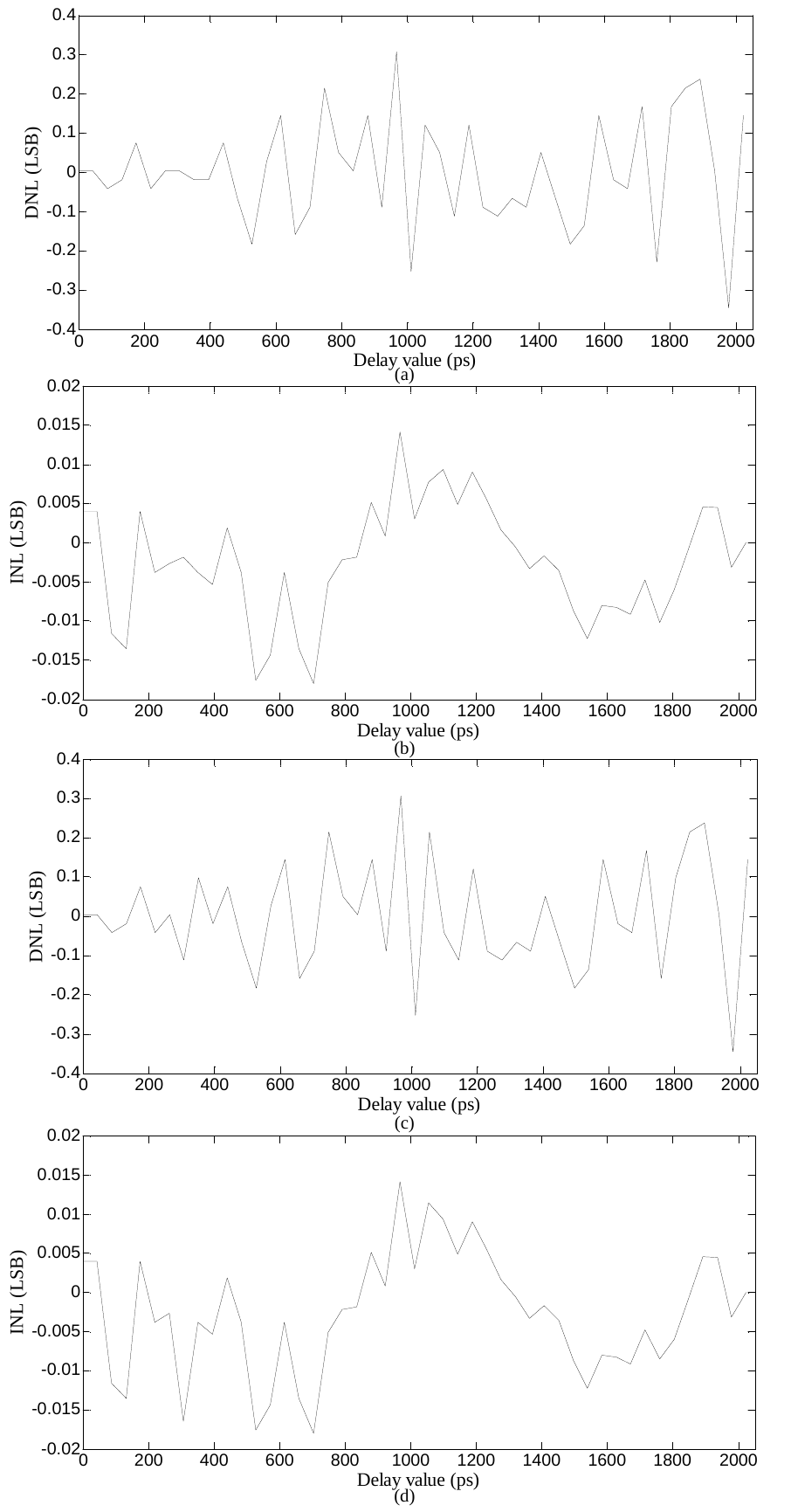}

\caption{DNLs and INLs under different voltage conditions. (a) DNL with 1050
mV; (b) INL with 1050 mV; (c) DNL with 1150 mV; (d) INL with 1150
mV.}
\end{figure}

\begin{figure}[tbh]
\centering\includegraphics[width=1\columnwidth]{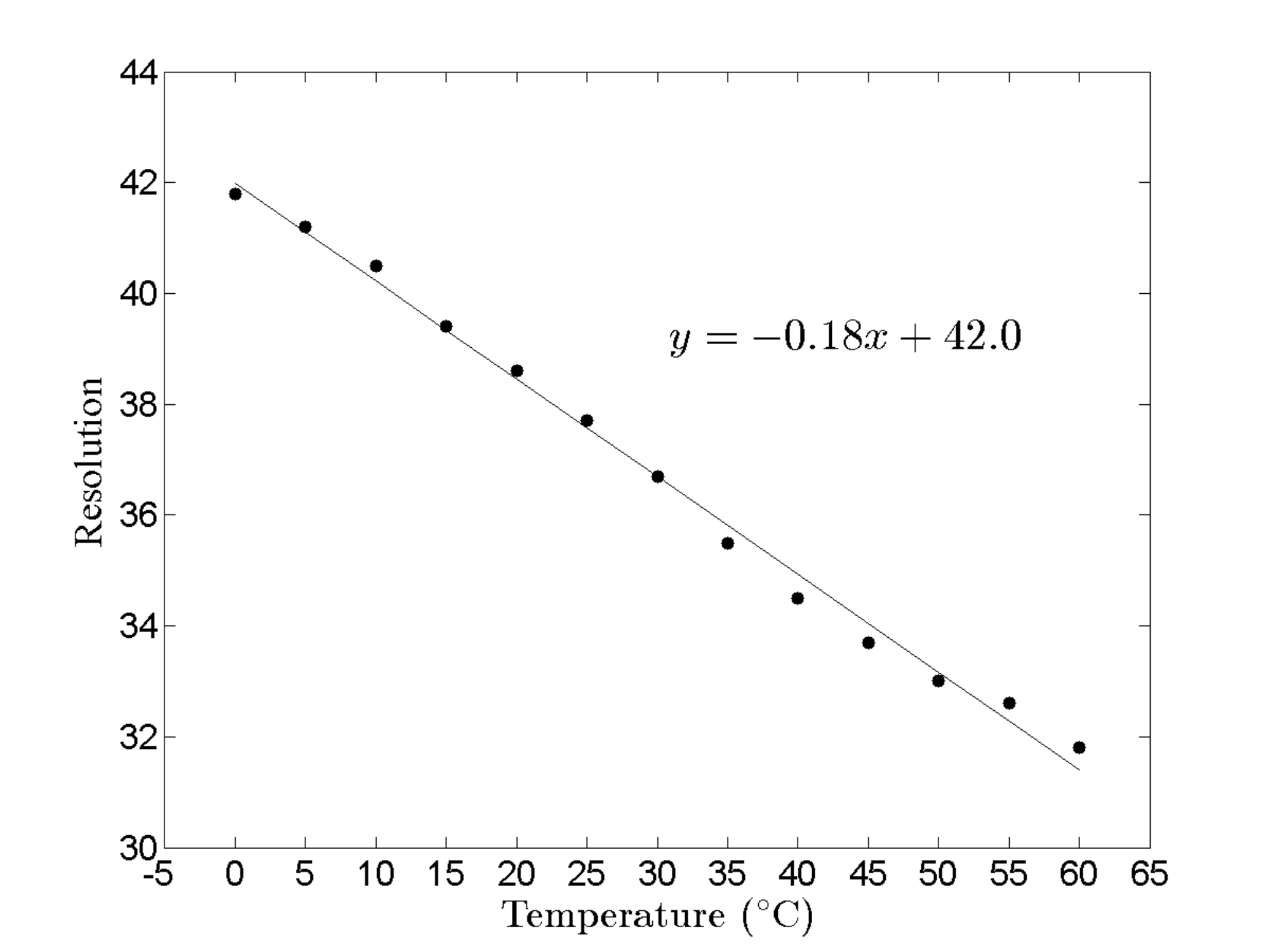}

\caption{Resolution versus temperature.}
\end{figure}

\begin{figure}[tbh]
\centering\includegraphics[width=0.9\columnwidth]{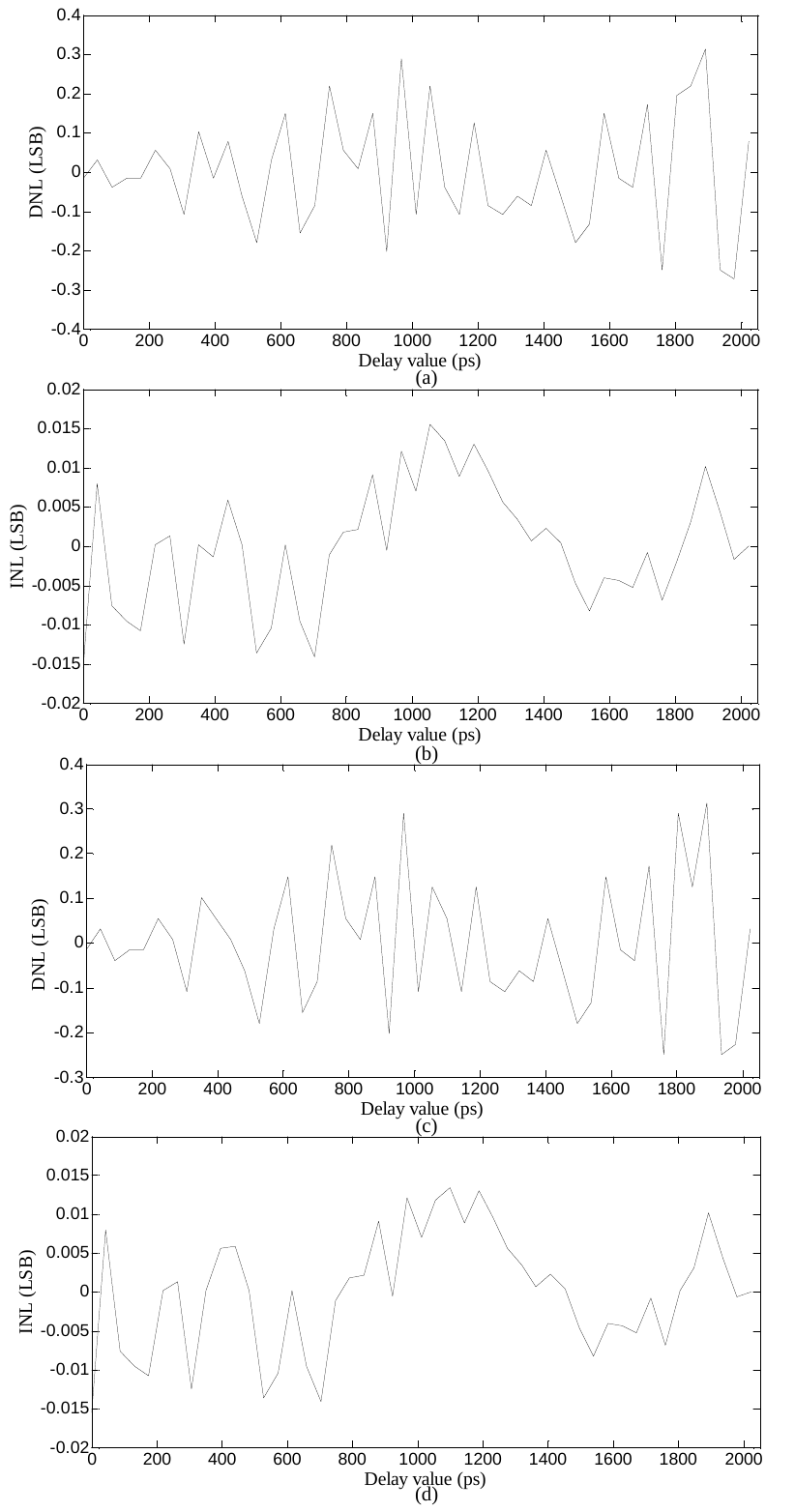}

\caption{DNLs and INLs under different temperature conditions. (a) DNL with
0 $^{\circ}$C; (b) INL with 0 $^{\circ}$C; (c) DNL with 60 $^{\circ}$C;
(d) INL with 60 $^{\circ}$C.}
\end{figure}

The proposed delay generator was implemented in the EP3SE110F1152I3
Stratix III FPGA from Altera on a self-designed test board. The board
also contains a USB interface fulfilled by the CY7C68013 chip from
which the digitally programmable delay values $(m,\thinspace n)$
are received. The delay generator is configured in a special testing
mode for the measurement convenience. In this mode, the delay generator
continues to function automatically every 1 $\mu$s by self-reset.
The automation procedure is controlled by the USB interface. The two
delay outputs are brought out through two SMA connectors and prolonged
by two coaxial cables of the same length to the external 740Zi Lecroy
digital oscilloscope with 4 GHz signal bandwidth and 40 GS/s real
time sample rate. The oscilloscope directly provides a delay difference
measurement tool which calculates the statistical parameters such
as the mean and the RMS jitter. More than 5000 hits are accumulated
to calculate the statistical results for every testing point.

Fig.5 depicts the DNL and INL results which cover the coarse clock
period of 2000 ps under the nominal supply voltage of 1100 mV and
environmental temperature of 20 $^{\circ}$C. The maximal fine delay
number $n$ in the figure is 52 and corresponds to the generated fine
delay of about 2010 ps, so the resolution is 2010/52=38.6 ps. The
DNL\,\textbackslash{}\,INL is in the range of (-0.18 LSB, 0.24 LSB)\,\textbackslash{}\,(-0.02
LSB, 0.01 LSB) according to Fig.5. The RMS jitter is tested and depicted
in Fig.4 which is used to decide the lower bound of the resolution.
The biggest tested RMS jitter for our delay generator is 33.6 ps when
$n$ equals the maximal value of 52 for the fine dynamic working range.
Larger delay value will cause $n$ to recirculate from 0 which is
constrained never to exceed 52 by means of increasing the coarse delay
number $m$ accordingly. 

The coarse delay generator uses the clock signal compensated by the
PLL which gives little contribution to the overall RMS jitter. Fig.6
shows the waveform and histogram captured by the Lecroy digital oscilloscope
when the programmed delay numbers are set to $(m,\thinspace n)$=(0,\,6)\,\textbackslash{}\,(3,\,6)
corresponding to the expected delay values of $0\times2000+6\times38.6=232$
ps\,\textbackslash{}\,$3\times2000+6\times38.8=6232$ ps. The tested
delay intervals are 228 ps\,\textbackslash{}\,6227 ps and the tested
RMS jitters are 6.9 ps\,\textbackslash{}\,5.7 ps. It shows they
obtain nearly the same RMS jitter and manifests the fact that $n$
is the decisive parameter contributing to the RMS. According to the
working principle, the proposed delay generator has a maximal dead
time of $T_{s}\cdot n_{max}=4\thinspace\textrm{ns}\cdot52=208\thinspace\textrm{ns}.$

Finally the delay stability against environmental changes such as
voltage and temperature is evaluated. For voltage stability consideration,
the core voltage of the FPGA chip is scanned from 1050 mV to 1150
mV with a 10 mV interval. The obtained resolution versus voltage is
depicted and fitted using a linear function in Fig.7. The fitted voltage
drifting coefficient is -0.465 ps/mV according to Fig.7. To visualize
the influence of the voltage drift to the DNL and INL performance,
we depict the corresponding DNLs and INLs under the voltage conditions
of 1050 mV and 1150 mV respectively in Fig.8. It can be seen that
voltage drift has very little influence to the DNLs and INLs of which
the DNLs lie in the range of (-0.4 LSB, 0.4 LSB) and the INLs in the
range of (-0.02 LSB, 0.02 LSB).

For temperature stability consideration, the entire test board implementing
the delay generator is placed in a high-low temperature test chamber
by using which a temperature dynamic range from 0 $^{\circ}$C to
60 $^{\circ}$C with a 5 $^{\circ}$C interval is covered. The obtained
resolution versus temperature is depicted and fitted using a linear
function in Fig.9. The fitted temperature drifting coefficient is
-0.18 ps/$^{\circ}$C according to Fig.9. Although the temperature
coefficient for a single delay line should be positive as in \cite{RGiordano2015},
a negative coefficient is still possible for the Vernier delay line
since the temperature change may have much larger influence to the
fast delay line compared with the slow delay line. Finally the corresponding
DNLs and INLs under the temperature conditions of 0 $^{\circ}$C and
60 $^{\circ}$C respectively are depicted in Fig.8. Similarly we conclude
that temperature drift has very little influence to the DNLs and INLs
of which the DNLs lie in the range of (-0.4 LSB, 0.4 LSB) and the
INLs in the range of (-0.02 LSB, 0.02 LSB).

The proposed delay generator is rather resource cost efficient. The
compilation report shows the used LUT occupation percentage is 668/85200
(0.7\%) and the used register occupation percentage is 146/85200 (0.2\%).
All required resources are ordinary and widely exist in the FPGA fabric,
and this characterization improves its usage convenience and releases
the constructing demands.

\section{Conclusions}

We present an delay generator architecture working in the Vernier
type based on the dedicated carry chains in the FPGA. The delay generator
was implemented in the EP3SE110F1152I3 Stratix III device from Altera
on a self-designed test board. Test result shows the obtained resolution
is 38.6 ps, and the DNL\,\textbackslash{}\,INL is in the range of
(-0.18 LSB, 0.24 LSB)\,\textbackslash{}\,(-0.02 LSB, 0.01 LSB) under
the nominal supply voltage of 1100 mV and environmental temperature
of 20 $^{\circ}$C. The delay generator is rather resource cost efficient
which uses only 668 LUTs and 146 registers in total. We believe the
proposed delay generator can present powerful usefulness in many physical
instruments due to its high-resolution and structural simplicity. 
\begin{acknowledgments}
This work was supported by the Fundamental Research Funds for the
Central Universities under Grants 30916014112-019 and 30916011349.
\end{acknowledgments}


\begin{thebibliography}{10}
	\providecommand{\url}[1]{#1}
	\csname url@samestyle\endcsname
	\providecommand{\newblock}{\relax}
	\providecommand{\bibinfo}[2]{#2}
	\providecommand{\BIBentrySTDinterwordspacing}{\spaceskip=0pt\relax}
	\providecommand{\BIBentryALTinterwordstretchfactor}{4}
	\providecommand{\BIBentryALTinterwordspacing}{\spaceskip=\fontdimen2\font plus
		\BIBentryALTinterwordstretchfactor\fontdimen3\font minus
		\fontdimen4\font\relax}
	\providecommand{\BIBforeignlanguage}[2]{{%
			\expandafter\ifx\csname l@#1\endcsname\relax
			\typeout{** WARNING: IEEEtran.bst: No hyphenation pattern has been}%
			\typeout{** loaded for the language `#1'. Using the pattern for}%
			\typeout{** the default language instead.}%
			\else
			\language=\csname l@#1\endcsname
			\fi
			#2}}
	\providecommand{\BIBdecl}{\relax}
	\BIBdecl
	
	\bibitem{YLiu2003}
	Y.~Liu, H.~Li, Y.~Wang, T.~Xing, H.~Baghaei, J.~Uribe, R.~Farrell, and W.-H.
	Wong, ``A programmable high-resolution ultra-fast delay generator,''
	\emph{IEEE Transactions on Nuclear Science}, vol.~50, no.~5, pp. 1487--1490,
	Oct 2003.
	
	\bibitem{JPMartin2000}
	J.~P. Martin and K.~Ragan, ``A programmable nanosecond digital delay and
	trigger system,'' in \emph{Nuclear Science Symposium Conference Record, 2000
		IEEE}, vol.~2, 2000, pp. 12/141--12/144 vol.2.
	
	\bibitem{RGinosar2011}
	R.~Ginosar, ``Metastability and synchronizers: A tutorial,'' \emph{IEEE Design
		Test of Computers}, vol.~28, no.~5, pp. 23--35, Sept 2011.
	
	\bibitem{HUJang-RSI}
	H.~U. Jang, J.~Blieck, G.~Veshapidze, M.~L. Trachy, and B.~D. DePaola, ``An
	auto-incrementing nanosecond delay circuit,'' \emph{Review of Scientific
		Instruments}, vol.~78, no.~9, p. 094702, 2007.
	
	\bibitem{JKalisz-RSI}
	J.~Kalisz, A.~Poniecki, and K.~Różyc, ``A simple, precise, and low jitter
	delay/gate generator,'' \emph{Review of Scientific Instruments}, vol.~74,
	no.~7, pp. 3507--3509, 2003.
	
	\bibitem{YSong2011}
	Y.~Song, H.~Liang, L.~Zhou, J.~Du, J.~Ma, and Z.~Yue, ``Large dynamic range
	high resolution digital delay generator based on {FPGA},'' in
	\emph{Electronics, Communications and Control (ICECC), 2011 International
		Conference on}, Sept 2011, pp. 2116--2118.
	
	\bibitem{RGiordano2015}
	R.~Giordano, F.~Ameli, P.~Bifulco, V.~Bocci, S.~Cadeddu, V.~Izzo, A.~Lai,
	S.~Mastroianni, and A.~Aloisio, ``High-resolution synthesizable
	digitally-controlled delay lines,'' \emph{IEEE Transactions on Nuclear
		Science}, vol.~62, no.~6, pp. 3163--3171, Dec 2015.
	
	\bibitem{KeCui_TDC2017}
	K.~Cui, Z.~Ren, X.~Li, Z.~Liu, and R.~Zhu, ``A high-linearity,
	ring-oscillator-based, {Vernier} time-to-digital converter utilizing carry
	chains in {FPGAs},'' \emph{IEEE Transactions on Nuclear Science}, vol.~64,
	no.~1, pp. 697--704, Jan 2017.
	
	\bibitem{JYWon_TDC2016}
	J.~Y. Won and J.~S. Lee, ``Time-to-digital converter using a tuned-delay line
	evaluated in 28-, 40-, and 45-nm {FPGAs},'' \emph{IEEE Transactions on
		Instrumentation and Measurement}, vol.~65, no.~7, pp. 1678--1689, July 2016.
	
	\bibitem{WangYG_TDC2016}
	Y.~Wang and C.~Liu, ``A 3.9 ps time-interval {RMS} precision time-to-digital
	converter using a dual-sampling method in an {UltraScale FPGA},'' \emph{IEEE
		Transactions on Nuclear Science}, vol.~63, no.~5, pp. 2617--2621, Oct 2016.
	
	\bibitem{JYu2010}
	J.~Yu, F.~F. Dai, and R.~C. Jaeger, ``A 12-bit {V}ernier ring time-to-digital
	converter in 0.13 um {CMOS} technology,'' \emph{IEEE Journal of Solid-State
		Circuits}, vol.~45, no.~4, pp. 830--842, April 2010.
	
	\bibitem{JKalisz_TDC1997}
	J.~Kalisz, R.~Szplet, J.~Pasierbinski, and A.~Poniecki,
	``Field-programmable-gate-array-based time-to-digital converter with 200-ps
	resolution,'' \emph{IEEE Transactions on Instrumentation and Measurement},
	vol.~46, no.~1, pp. 51--55, Feb 1997.
	
	\bibitem{HaiWang_TDC2013}
	H.~Wang, M.~Zhang, and Q.~Yao, ``A new realization of time-to-digital
	converters based on {FPGA} internal routing resources,'' \emph{IEEE
		Transactions on Ultrasonics, Ferroelectrics, and Frequency Control}, vol.~60,
	no.~9, pp. 1787--1795, Sep 2013.
	
	\bibitem{AAAbidi2006}
	A.~A. Abidi, ``Phase noise and jitter in {CMOS} ring oscillators,'' \emph{IEEE
		Journal of Solid-State Circuits}, vol.~41, no.~8, pp. 1803--1816, Aug 2006.
	
\end{thebibliography}
\end{document}